\documentclass[10pt]{article}
\usepackage{amsmath}
\usepackage{amsfonts}
\usepackage{setspace}

\usepackage{amsmath,amssymb}
\usepackage{graphicx}
\usepackage{color}
\usepackage{url}
\usepackage{tabularx} 
\usepackage{ragged2e} 
\usepackage{booktabs} 

\usepackage{amsmath}
\usepackage{setspace}
\usepackage{array,caption}
\usepackage[labelfont=bf]{caption}
\usepackage{longtable}
\usepackage{booktabs}

\newtheorem{theorem}{Theorem}
\newtheorem{lemma}{Lemma}
\newtheorem{corollary}{Corollary}

\newtheorem{proposition}{Proposition}
\newtheorem{remark}{Remark}

\newtheorem{example}{Example}

\newcommand{\RNum}[1]{\uppercase\expandafter{\romannumeral #1\relax}}

\newcommand{\F}{\ensuremath{\mathbb F}}

\newcommand{\done}{\hfill $\Box$ }

\newcommand{\notequiv}{{\,\not\equiv\, }}

\newcommand{\fp}{{\mathbb F}_{p}}

\newcommand{\ls}[1]
    {\dimen0=\fontdimen6\the\font\lineskip=#1\dimen0
     \advance\lineskip.5\fontdimen5\the\font
     \advance\lineskip-\dimen0
     \lineskiplimit=0.9\lineskip
     \baselineskip=\lineskip
     \advance\baselineskip\dimen0
     \normallineskip\lineskip\normallineskiplimit\lineskiplimit
     \normalbaselineskip\baselineskip
     \ignorespaces}

\begin{document}

\bibliographystyle{abbrv}

\title{Several classes of optimal $p$-ary cyclic codes
with minimal distance four}

\author{
Gaofei Wu
\thanks{G. Wu and H. Liu are with the State Key Laboratory of Integrated Service Networks, School of Cyber Engineering, Xidian University, Xi'an, 710071, China.
Email: gfwu@xidian.edu.cn, 2473116010@qq.com.}
, Huan Liu
and Yuqing Zhang
\thanks{Y. Zhang is with National Computer Network Intrusion Protection Center, University of Chinese Academy of Sciences, Beijing 101408, China. Email: zhangyq@ucas.ac.cn.}
}
\date{}
\maketitle

\thispagestyle{plain} \setcounter{page}{1}
\ls{1.5}
\begin{abstract}

Cyclic codes are a subclass of linear codes and
have wide applications in data storage systems, communication systems and consumer electronics
due to their efficient encoding and decoding algorithms.
Let $p\ge 5$ be an odd prime and $m$ be a positive integer.
Let $\mathcal{C}_{(1,e,s)}$ denote the $p$-ary cyclic code with three nonzeros
$\alpha$, $\alpha^e$, and $\alpha^s$, where $\alpha $ is a generator of $\F_{p^m}^*$,
 $s=\frac{p^m-1}{2}$, and $2\le e\le p^m-2$.
In this paper, we present four classes of optimal $p$-ary  cyclic codes $\mathcal{C}_{(1,e,s)}$
with parameters $[p^m-1,p^m-2m-2,4]$ by analyzing
the solutions of certain polynomials over finite fields.
Some previous results about optimal quinary cyclic codes with parameters
$[5^m-1,5^m-2m-2,4]$ are special cases of our constructions.
In addition,
by       analyzing the irreducible factors of certain polynomials over $\F_{5^m}$,
we  present two classes of
      optimal quinary cyclic codes  $\mathcal{C}_{(1,e,s)}$.
\end{abstract}


\section{Introduction}

Cyclic codes are a very important subclass of linear codes,
and have applications in consumer electronics, data storage systems and
communication systems due to their efficient encoding and
decoding algorithms. Throughout this paper,
Let $p$ be a prime and  $m$ be a positive integer. Let ${{\mathbb{F}}_{{{p}^{m}}}}$ denote the  finite field with ${{p}^{m}}$ elements.
An $[n,k,d]$ linear code over $\fp$ is a $k$-dimensional subspace of $\fp^n$ with minimum Hamming distance $d$.
An $[n,k]$ \emph{cyclic} code $ \mathcal{C}$ is an $[n,k]$ linear code with the property that any cyclic shift of a codeword is another codeword of $\mathcal{C}$.
Let $\gcd(n,p)=1$. By identifying any codeword $(c_0,c_1,\cdots, c_{n-1})\in \mathcal{C}$ with
$$
c_0+c_1x+c_2x^2+\cdots+c_{n-1}x^{n-1}\in \fp[x]/(x^n-1),
$$
any cyclic code of length $n$ over $\fp$ corresponds to an ideal of the polynomial ring $\fp[x]/(x^n-1)$.
Notice  that every ideal of $\fp[x]/(x^n-1)$ is principal. Thus, any cyclic code can be expressed as
$\langle g(x)\rangle$, where $g(x)$ is monic and has the least degree. The polynomial $g(x)$ is called the \emph{generator polynomial}
and $h(x)=(x^n-1)/g(x)$ is called the \emph{parity-check polynomial} of $\mathcal{C}$.
 Let $\alpha$ be a generator   of  ${{\mathbb{F}}_{{{p}^{m}}}^*}$ and
 let ${{m}_{\alpha^{i}}}(x)$ denote the minimal polynomial of ${{\alpha }^{i}}$ over  ${{\mathbb{F}}_{p}}$.
We denote by $\mathcal{C}_{(i_1,i_2,\cdots, i_t)}$
the  cyclic code with generator polynomial ${{m}_{{{\alpha}^{{{i}_{1}}}}}}(x){{m}_{{{\alpha}^{{{i}_{2}}}}}}(x)\cdots{{m}_{{{\alpha}^{{{i}_{t}}}}}}(x)$.
 In 2005, Carlet, Ding, and Yuan  \cite{carletdingit05}
 constructed some
 optimal ternary cyclic codes $ \mathcal{C}_{(1,e)}$ with parameters $[3^m-1,3^m-2m-1,4]$ by using perfect
 nonlinear monomials $x^e$.
   In 2013,
Ding and Helleseth
 \cite{Ding-Helleseth} constructed  several
 classes of optimal ternary cyclic codes $\mathcal{C}_{(1,e)}$ by  using monomials including  almost
  perfect   nonlinear  (APN)  monomials.
Moreover, they  presented  nine open problems on optimal  ternary cyclic codes $\mathcal{C}_{(1,e)}$.
Two of them were solved in
 \cite{lizhou15} and \cite{liffa14}. 
In \cite{liffa14}, Li et al.  also presented several classes
of optimal ternary cyclic codes with parameters $[3^m-1,3^m-2m-1,4]$ or
$[3^m-1,3^m-2m-2,5]$.
Another open problem presented in \cite{Ding-Helleseth} were settled separately by     \cite{hanyanffa19}  and \cite{liucaoludcc2020}.
  In 2016, Wang and Wu \cite{wangwu16} presented
  four classes of optimal
  ternary cyclic codes with parameters
  $[3^m-1,3^m-2m-1,4]$ by analyzing the solutions
  of certain equations over $\F_{3^m}$. It was shown that
  some previous results about optimal ternary cyclic codes given in \cite{Ding-Helleseth}\cite{dinggaozhouit13}\cite{liffa14}\cite{zhoudingitc13}
  are special cases of the constructions given in \cite{wangwu16}.
There are some other  optimal ternary cyclic codes were constructed in the literature, see
 \cite{fanlizhouffa16}\cite{liucao21}\cite{yanzhouduffa18}\cite{zhaoluoffa22}\cite{zhahuffa2021}\cite{zhahu20} and the references therein.

However, there are only a few constructions of optimal $p$-ary $(p\ge 5)$
cyclic codes.
  In 2016, Xu and Cao \cite{xucaoccds16} presented
  some  optimal $p$-ary  cyclic codes  $\mathcal{C}_{(0,1,e)}$ with  generator polynomial
      $(x-1){{m}_{\alpha}}(x){{m}_{{{\alpha }^{e}}}}(x)$
  by using perfect nonlinear monomials. They also constructed some
  optimal quinary cyclic codes with parameters $[5^m-1,5^m-2m-2,4]$.
  In 2020, Fan and Zhang \cite{fanzhagn20}
  proposed a sufficient and necessary condition
  on $e$
  for quinary cyclic codes  $\mathcal{C}_{(1,e,s)}$ with  generator polynomial
      $(x+1){{m}_{\alpha}}(x){{m}_{{{\alpha }^{e}}}}(x)$ to be optimal,   where $s=\frac{p^m-1}{2}$ and
   $x+1$ is the minimal polynomial of $\alpha^s$ over $\F_p$. Several optimal quinary cyclic codes
    $\mathcal{C}_{(1,e,s)}$ with parameters $[5^m-1,5^m-2m-2,4]$ were also presented in
    \cite{fanzhagn20}. Recently, Liu and Cao \cite{liucao20} gave four classes of
     optimal quinary cyclic codes  $\mathcal{C}_{(0,1,e)}$   and  $\mathcal{C}_{(1,e,s)}$.
     We recall the known optimal quinary  cyclic codes $\mathcal{C}_{(1,e,s)}$ in Table \ref{table: test}.

        In this paper, we will focus on constructions of optimal $p$-ary  cyclic codes
     $\mathcal{C}_{(1,e,s)}$ with  generator polynomial
      $(x+1){{m}_{\alpha}}(x){{m}_{{{\alpha }^{e}}}}(x)$.
      As far as we know,  
      there are no  constructions of 
      optimal $p$-ary (for general $p\ge 5$) 
cyclic codes $\mathcal{C}_{(1,e,s)}$. 
    By analyzing
      the solutions of certain polynomials over finite fields,
          we  present four classes of optimal $p$-ary ($p\geq 5$) cyclic codes $\mathcal{C}_{(1,e,s)}$
      with parameters $[p^m-1,p^m-2-2m,4]$.
      We will show that some previous
      results on optimal  quinary cyclic codes are special cases
      of our constructions.
      In addition, we also present two  classes of
      optimal quinary cyclic codes  $\mathcal{C}_{(1,e,s)}$ by
       analyzing the irreducible factors of certain polynomials over $\F_{5^m}$.

The rest of this paper is organized
as follows. In Section 2, we introduce some preliminaries.
Four  classes of optimal $p$-ary ($p\geq 5$) cyclic codes $\mathcal{C}_{(1,e,s)}$ are given in Section 3.
In Section 4, we present two  classes  of
      optimal quinary cyclic codes. Concluding  remarks and some open problems are given in Section 5.

\newcommand{\tabincell}[2]{\begin{tabular}{@{}#1@{}}#2\end{tabular}}
\begin{center}
	\setlength\LTleft{0pt}
	\setlength\LTright{0pt}
	\scriptsize
	\begin{longtable}{@{\extracolsep{\fill}}lll@{}}
		\caption {Known cyclic codes ${\mathcal{C}_{(1,e,s)}}$ with parameters $[{{5}^{m}}-1,{{5}^{m}}-2m-2,4 ]$ }
		\label{table: test}\\
		\endfirsthead
		Table \ref{table: test} \\\toprule
		\endhead
		\bottomrule
		\endfoot
		\bottomrule
		\endlastfoot
		\toprule
		Values of $e$ or requirements on $e$ & Conditions  & Ref. \\
		\hline
		
		$e=\frac{{{5}^{m}}-1}{2}+\frac{{{5}^{t}}+1}{2}$ & $t$ is even, $\gcd (m,t)=1$. & \cite[Cor. 13]{liucao20} \\
		$e=\frac{{{5}^{t}}+1}{2}$ & $t$ is odd, $m$ is odd,  $\gcd (m,t)=1$.  & \cite[Cor. 11]{liucao20} \\
		\tabincell{l}{$e({{5}^{k}}+1)\equiv {{5}^{t}}+1(\bmod {{5}^{m}}-1)$, \\ and $e\equiv 3(\bmod 4)$} & \tabincell{l} {$\gcd (m,t\pm k)=1$, $m$ is  odd. }  & \cite[Thm. 10]{liucao20} \\
		\tabincell{l}{$e({{5}^{k}}-1)\equiv {{5}^{t}}-1(\bmod {{5}^{m}}-1)$,\\ and $e\equiv 0(\bmod 4)$ or $e\equiv 3(\bmod 4)$} & $\gcd (m,t)=\gcd (m,k)=\gcd (m,t\pm k)=1$, $m$ is odd.  &  \cite[Thm. 19]{liucao20} \\
		$e=\frac{{{5}^{t}}-1}{4}$ & \tabincell{l}{ $\gcd (m,t)=\gcd(m,t\pm1)=1$, $t\equiv 3(\bmod 4)$  } &  \cite[Cor. 20]{liucao20} \\
		$e=\frac{{{5}^{m}}-1}{4}+\frac{{{5}^{t}}-1}{4}$ & \tabincell{l}{$\gcd (m,t)=\gcd(m,t\pm1)=1$, $t\equiv 3(\bmod 4)$, $m\equiv 1(\bmod 4)$.  }  &  \cite[Cor. 21]{liucao20} \\
		
		\tabincell{l}{$({{5}^{h}}-1)e\equiv {{5}^{t}}-{{5}^{k}}(\bmod {{5}^{m}}-1)$} & see Ref. \cite{fanzhagn20} & \cite[Thm. 1]{fanzhagn20} \\
	
		\tabincell{l}{$({{5}^{h}}+1)e\equiv {{5}^{t}}+{{5}^{k}}(\bmod {{5}^{m}}-1)$} & see Ref. \cite{fanzhagn20} &\cite[Thm. 2]{fanzhagn20} \\
	
		$e=\frac{{{5}^{m}}-1}{2}+{{5}^{h}}+1$ &  \tabincell{l} {$0\le h<m$, $h\ne \frac{m}{2}$ if $m$ is even. }  &\cite[Thm. 3]{fanzhagn20} \\
		$e={{5}^{h}}+1$ & \tabincell{l}{$m\equiv 0(\bmod 4)$,  $h=0$, \\ or $\gcd (h,m)=1$,  or $\frac{m}{\gcd (h,m)}$ is odd. }&\cite[Thm. 5]{fanzhagn20} \\
	
		$e={{5}^{h}}+2$ & \tabincell{l}{$m$ is odd;  $m\equiv 0(\bmod 4)$,  $h=0,\frac{m}{2}$; \\$m\equiv 0(\bmod 4)$,$h=0$ or $\gcd(h,m)=1,2$ }&\cite[Thm. 7]{fanzhagn20} \\
	
		$e={{5}^{h+1}}+{{5}^{h}}+1$ & see Ref. \cite{fanzhagn20} &\cite[Thm. 8]{fanzhagn20} \\
	
		$e=h({{5}^{m-1}}-1)(\bmod {{5}^{m}}-1)$ &see Ref. \cite{fanzhagn20}  &\cite[Thm. 9]{fanzhagn20} \\
	
			$e=\frac{{{5}^{m}}-1}{2}+h$  &see Ref. \cite{fanzhagn20}
			&\cite[Thm. 10]{fanzhagn20} \\
			$e={{5}^{m}}-{{5}^{m-h}}+3$	& $h=0$; $h=1$ with $m \notequiv 0 \pmod9$ and $m \notequiv 0 \pmod8$ &Theorem  \ref{45h-1thm} \\
			$e=\frac{{{5}^{m}}-1}{2}-3$ &	$m$ is odd  & Theorem  \ref{p5-3thm}\\
			$e=\frac{{{5}^{m}}-1}{2}+\frac{5^h+1}{2}$&	$m$ is even and $\gcd(h,m)=1$     &	Theorem \ref{p5h+1div2thm}\\

			\bottomrule
		\end{longtable}
	\end{center}


\section{Preliminaries}

Let $p$ be a prime and $m$ be a positive integer.
The $p$-cyclotomic coset modulo $p^m-1$ containing $j$ is defined as
\[C_j=\{j\cdot p^r\pmod{p^m-1}: r=0,1,\cdots,l_j-1\},\]
where $l_j$ is the least positive integer such that $j\cdot p^{l_j} \equiv j \pmod{p^m-1}.$
Thus the size of $C_j$ is $|C_j|=l_j$. It is known that $l_j|m$. The smallest integer
in $C_j$ is called the coset leader of $C_j$.

\begin{lemma}\cite[Lemma 1 and Lemma 2]{xucaoccds16} \label{ce}
	 Let $p$ be a prime and $m$ be a positive
 integer. Let $n={{p}^{m}}-1$.  For any $1\le e\le n-1$,  if
 one of the following conditions is satisfied, then $l_e=|{C_{e}} |=m$:
 \begin{itemize}
   \item [1)]  $1\le \gcd (e, n)\le p-1$,
   \item  [2)]  $ \gcd (e, n)\gcd(p^j-1,n) \notequiv 0\pmod n$
    for all $1\le j<m$.
 \end{itemize}
\end{lemma}

The following lemma shows that
the $p$-ary ($p\ge 5$) cyclic code
  ${\mathcal{C}_{(1,e,s)}}$ is optimal if it has parameters
  $[p^m-1,p^m-2m-2,4]$.

\begin{lemma} \cite[Lemma 5]{xucaoccds16} \label{d=4}
Let $p\ge 5$ be an odd prime and   $m>1$. For any $e\notin {{C}_{1}}$ and
 $\left| {C_{e}} \right|=m$,  the minimum distance d of the cyclic code
  ${\mathcal{C}_{(1,e,s)}}$ over $\mathbb{F}_p$ satisfies $d\le 4$.
	
\end{lemma}

\begin{lemma}\cite[Theorem 3.20]{lidl83}\label{irrdivide}
 For every finite field $\mathbb{F}_{p^m}$ and every positive integer $r$,
 the product of all monic  irreducible polynomials over $\mathbb{F}_{p^m}$ whose degrees divide $r$ is equal to ${x}^{(p^m)^{r}}-x$.
\end{lemma}

\begin{lemma}\cite[Theorem 2.14]{lidl83}   \label{root}
 Let  $f(x)$ be an irreducible
polynomial over $\F_{p^m}$ of degree $r$.
Then $f(x) = 0$ has a root $x$ in $\F_{p^{mr}}$. Furthermore, all the
roots of $f(x) = 0$ are simple and are given by the $r$  distinct elements $x, x^{p^m}, x^{p^{2m}}, \cdots ,
x^{p^{m(r-1)}}$ of $\F_{p^{mr}}$.
\end{lemma}

For an odd prime $p$ and  $i,j \in\{-1,1\}$,
define the set $$N_{i,j}=\{x| x^{\frac{p^m-1}{2}}=i,
 (x+1)^{\frac{p^m-1}{2}}=j,x\in\F_{p^m}\setminus \{0,-1\} \}.$$
Let  $|N_{i,j}|$ be the number of elements in $N_{i,j}$.

\begin{lemma}\cite[Lemma 6]{storer67}\label{numofso}
  If $p^m\equiv 1\pmod 4, $ then $|N_{1,1}|=
  \frac{p^m-5}{4}$, $|N_{1,-1}|=|N_{-1,1}|=|N_{-1,-1}|=\frac{p^m-1}{4}$.
   If $p^m\equiv 3\pmod 4, $ then
   $|N_{1,1}|=|N_{-1,1}|=|N_{-1,-1}|=\frac{p^m-3}{4}$,
   $|N_{1,-1}|=
  \frac{p^m+1}{4}$.
\end{lemma}


\section{Four   classes of  optimal $p$-ary cyclic codes}
In this section,
we will
 propose four classes of optimal $p$-ary cyclic codes ${\mathcal{C}_{(1,e,s)}}$ with parameters $[{{p}^{m}}-1,{{p}^{m}}-2m-2,4]$
 by analyzing the solutions of certain equations over $\F_{p^m}$.
     Some quinary cyclic codes given in \cite{fanzhagn20}\cite{tianzhang17} are  special cases of our constructions.
    Throughout this section, let
    $p\ge 5$ be  an odd prime and $m$ be a  positive integer.
     We
    denote by  $\eta(x)=x^{\frac{p^m-1}{2}}$ the  quadratic character on $\F_{p^m}$.
%

\subsection{The first class of optimal $p$-ary cyclic  codes with minimal distance 4}

Let $e={{p}^{m}}-2$.
In this subsection, we will show that
 ${\mathcal{C}_{(1,e,s)}}$ is an  optimal $p$-ary cyclic code with parameters $[{{p}^{m}}-1,{{p}^{m}}-2m-2,4]$
 if    ${{p}^{m}}\equiv 1\pmod4$.

\begin{theorem}\label{pm-2thm}
	Let  $e={{p}^{m}}-2$ and  $s=\frac{{{p}^{m}} -1}{2}$,
  where $p\geq 5$ is an odd prime.  If  ${{p}^{m}}\equiv 1\pmod4$,  then ${\mathcal{C}_{(1,e,s)}}$ is an  optimal $p$-ary cyclic code with parameters $[{{p}^{m}}-1,{{p}^{m}}-2m-2,4]$.
\end{theorem}

{\em Proof:}
We first show that $e\notin {{C}_{1}}$ and $|C_e|=m$.
 If $e\in {{C}_{1}}$, then  ${{p}^{i}}e\equiv 1\pmod{ p^m-1} $, i.e., $p^m-1|{{p}^{i}}e-1$. As a result,
 $e\equiv 1 \pmod{ p-1}$, this is contrary to $e=p^m-2\equiv -1 \pmod{ p-1} $, thus $e\notin C_1$.  By $\gcd(p^m-2,p^m-1)=1$, we have  $|C_e|=m$ due to Lemma  \ref{ce}.
 Thus  the dimension of the cyclic code ${\mathcal{C}_{(1,e,s)}}$ is ${{p}^{m}}-2m-2$.

 In the following we show that
 ${\mathcal{C}_{(1,e,s)}}$ does not have a codeword of Hamming weight less than 4.
 It is  obviously that the minimum distance of $\mathcal{C}_{(1,e,s)}$ cannot  be 1.
Suppose that   ${\mathcal{C}_{(1,e,s)}}$ has a codeword of Hamming weight 2, then  there exist two elements ${{c}_{1}}$ and
${{c}_{2}}\in {{\mathbb{F}}_{p}^*}$ and two distinct elements ${{x}_{1}} $ and
${{x}_{2}}\in {{\mathbb{F}}_{p^m}^*}$ such that
\begin{equation}\label{weight2equ}
 \left\{ \begin{array}{{l}}
	{{c}_{1}}{{x}_{1}}+{{c}_{2}}{{x}_{2}}=0  \\
	{{c}_{1}}x_{1}^{e}+{{c}_{2}}x_{2}^{e}=0  \\
	{{c}_{1}}x_{1}^{s}+{{c}_{2}}x_{2}^{s}=0.
\end{array} \right.
\end{equation}
 Since $s=\frac{{{p}^{m}}-1}{2}$,
  we have
  ${x}_{1}^{s}=\pm1$ and ${x}_{2}^{s}=\pm1$.
  If $({x}_{1}^{s},{x}_{2}^{s})=(1,1)$,
  we have ${c}_{1}=-{c}_{2}$ and ${x}_{1}={x}_{2}$ by (\ref{weight2equ}), which is contrary to ${x}_{1}\ne {x}_{2}$.
  If $({x}_{1}^{s},{x}_{2}^{s})=(1,-1)$, again  by (\ref{weight2equ}),
   we have  ${c}_{1}={c}_{2}$  and ${x}_{1}=-{x}_{2}$.
    Then $1={x}_{1}^{s}={(-{x}_{2})}^{s}={x}_{2}^{s} $ since $s=\frac{p^m-1}{2}$ is even, which is contrary to ${x}_{2}^{s}=-1$. Thus $\mathcal{C}_{(1,e,s)}$ does not have  a codeword of Hamming weight 2.

The code $\mathcal{C}_{(1,e,s)}$ has a codeword of Hamming weight 3 if and only if there exist three elements ${c}_{1},{c}_{2},{c}_{3}\in \mathbb{F}_{p}^*$ and three distinct elements ${x}_{1},{x}_{2,},{x}_{3}\in \mathbb{F}_{p^m}^*$ such that
\begin{equation}\label{pm-21}
	\left\{ \begin{array}{l}
		{c}_{1}{x}_{1}+{c}_{2}{x}_{2}+{c}_{3}{x}_{3}=0  \\
		{c}_{1}x_{1}^{e}+{c}_{2}x_{2}^{e}+{c}_{3}x_{3}^{e}=0  \\
		{c}_{1}x_{1}^{s}+{c}_{2}x_{2}^{s}+{c}_{3}x_{3}^{s}=0.
	\end{array} \right.
\end{equation}
 Without lose of generality,  let $c_1=1$. Then (\ref{pm-21}) becomes
\begin{equation}\label{weight3equ}
 \left\{\begin{array}{l}
	{x}_{1}+{c}_{2}{x}_{2}+{c}_{3}{x}_{3}=0  \\
	x_{1}^{e}+{c}_{2}x_{2}^{e}+{c}_{3}x_{3}^{e}=0  \\
	x_{1}^{s}+{c}_{2}x_{2}^{s}+{c}_{3}x_{3}^{s}=0.
\end{array} \right.
\end{equation}
Due to symmetry, it is sufficient to  consider the following two cases.

	 Case 1),
   $({x}_{1}^{s},{x}_{2}^{s},{x}_{3}^{s})=(1,1, 1)$.
  In this case,
  $ 1+{c}_{2}+{c}_{3}=0$.
      Note that $({x}_{1}^{e},{x}_{2}^{e},{x}_{3}^{e})= ({x}_{1}^{-1},{x}_{2}^{-1},{x}_{3}^{-1})$.
	From the first equation of (\ref{weight3equ}), we get  ${x}_{1}=-({c}_{2}{x}_{2}+{c}_{ 3}{x}_{3})$,  substitute it into the second equation of  (\ref{weight3equ}), we obtain
	\begin{equation}\label{pm-22}
		-{({c}_{2}{x}_{2}+{c}_{3}{x}_{3})}^{-1}+{c}_{2}{x}_{2}^{-1}+{c}_{3}{x}_{3}^{-1}=0. \end{equation}
	Multiplying both sides of (\ref{pm-22}) by ${c}_{2}{x}_{2}+{c}_{3}{x}_{3} $ will lead to
	\begin{equation*}		-1+{c}_{2}{x}_{2}^{-1}({c}_{2}{x}_{2}+{c}_{3}{x}_{3})+{c}_{3}{x}_{3}^{-1}({c}_{2}{x}_{2}+{c}_{3}{x}_{3})=0,
	\end{equation*}
i.e.,
	\begin{equation*}
		(-1)+{c}_{2}^{2}+{c}_{2}{c}_{3}({x}_{2}^{-1}{x}_{3}+{x}_{2}{x}_{3}^{-1})+{c}_{3}^{2}=0.
	\end{equation*}
	Let ${x}_{2}^{-1}{x}_{3}=y$, then $y\notin \{0,1\}$. We have
	\begin{equation*}
		{c}_{2}{c}_{3}(y+\frac{1}{y})+({c}_{2}^{2}+{c}_{3}^{2}-1)=0,
	\end{equation*}
which is equivalent to
	\begin{equation*}
		{y}^{2}+\frac{{c}_{2}^{2}+{c}_{3}^{2}-1}{{c}_{2}{c}_{3}}y+1=0.
	\end{equation*}
By
	$1+{c}_{2}+{c}_{3}=0$, we get
$\frac{{c}_{2}^{2}+{c}_{3}^{2}-1}{{c}_{2}{c}_{3}}=
\frac{{c}_{2}^{2}+{(1+{c}_{2})}^{2}-1}{{c}_{2}{c}_{3}}=-2$,  thus ${y}^{2}-2y+1=0$,
i.e.,   $y=1$,  which is contrary to $y\notin \{0,1\}$.

Case 2),
   $({x}_{1}^{s},{x}_{2}^{s},{x}_{3}^{s})=(1,1,-1)$.
  In this case,
  $ 1+{c}_{2}-{c}_{3}=0$.
Similar as the
 proof of  Case 1), let ${x}_{2}^{-1}{x}_{3}=y$, we have  ${y}^{2}+2y+1=0$, i.e.,  $y=-1$, thus  ${x}_{2}=-{x }_{3}$. As a result,  $1={x}_{2}^{s}={(-{x}_{3})}^{s}={x}_ {3}^{s}$ due to $s=
\frac{p^m-1}{2}$ is even,  a contradiction with  ${x}_{3}^{s}=-1$.

To sum up,   $\mathcal{C}_{(1,e,s)}$ does not have  a codeword of  Hamming weight 3.
This completes the proof.
\done

\begin{remark}
	In 2017, Tian  et al. \cite{tianzhang17} constructed a class of  optimal quinary  cyclic code $\mathcal{C}_{(1,e,s)}$ with parameters  $[{5}^{m}-1,{5}^ {m}-2m-2,4]$, where  $e={5}^{m}-2$ and $s=\frac{5^m-1}{2}$.  It is the
 special case $p=5$ of Theorem \ref{pm-2thm}.
\end{remark}


\begin{example}
	Let $p=7$ and  $m=4$.  Then  $e={7}^{4}-2=2399$. Let $\alpha $ be the generator  of   $\mathbb{F}_{{7}^{4}}^*$
with
${\alpha}^{4}+5 {\alpha}^{2}+4\alpha+3=0$. Then the  code $\mathcal{C}_{(1,e,s)}$
 has  parameters $[2400,2391,4]$ and
 generator polynomial  ${x}^{9}+{x}^ {7}+{x}^{6}+2{x}^{5}+2{x}^{4}+{x}^{3}+{x }^{2}+1$.
\end{example}

\begin{example}
	Let $p=11$ and   $m=2$. Then   $e={11}^{2}-2=119$.
Let $\alpha $ be the generator  of   $\mathbb{F}_{11^2}^*$
with  ${\alpha}^{2}+7\alpha +2=0. $
Then the  code $\mathcal{C}_{(1,e,s)}$
 has  parameters $[120,115,4]$ and
 generator polynomial
 ${x}^{5}+6{x}^{4}+10{x}^{3}+10{x}^{2}+6x+1$.
\end{example}

\subsection{The second  class of optimal $p$-ary cyclic  codes with minimal distance 4}

In this subsection, we will present
a class of optimal $p$-ary cyclic  codes by using the exponent $e$ of the form
$e=\frac{{p}^{m}-1}{2}+{p}^{h}+1$.

\begin{lemma}\label{ph+1lem}
	Let $p\ge 5$ be an odd prime and $m$ be a positive integer. Let  $e=\frac{{p}^{m}-1}{2}+{p}^{h}+1$ and
 $s=\frac{p^m-1}{2}$. Suppose that   $h\ne\frac{m}{2}$ if $m$ is even, then $|{C}_{e}|=m$ and  $e\notin {C}_{1}$.
\end{lemma}
{\em Proof:}
If $e\in {C}_{1}$, then $\frac{{p}^{m}-1}{2}+{p}^{h}+1\equiv {p}^{i}\pmod{ {p}^{m}-1}$ for some $0\le i\le m-1$.
 Thus, ${p}^{m}-1|\frac{{p}^{m}-1}{2}+{p}^{h}+1-{p}^{i}$. As a
  consequence,  $p-1|\frac{{p}^{m}+1}{2}$.  Note that  $\gcd({p}^{m}+1,p-1)=2$, i.e.,
  $\gcd(\frac{{p}^{m}+1}{2},\frac{p-1}{2})=1$.  
  This is contrary to 
$p-1|\frac{{p}^{m}+1}{2}$. 
   Thus $e\in {C}_{1}$.

   In the following we show that $|{C}_{e}|=m$. By Lemma \ref{ce},
 it is sufficient to show that  for any $1\le j\le m-1$, $\gcd (e,{p}^{m}-1)\gcd ({p}^{j}-1,{p}^{m}-1)\not\equiv 0\pmod{ {p}^{m}-1}$. Note that
  $\gcd (e,{p}^{m}-1)|\gcd (2e,{p}^{m}-1)=\gcd(2({p}^{h}+1),{p}^{m}-1)$.
It is known  that
$$\gcd({p}^{h}+1,{p}^{m}-1)=\left\{
\begin{array}{ll}
2,& \textup{if}\ \frac{m}{\gcd(m,h)}\  \textup{is odd},\\
p^{\gcd(m,h)}+1,& \textup{if}\ \frac{m}{\gcd(m,h)}\  \textup{is even}.
\end{array}
\right.$$
  If $\frac{m}{\gcd (m,h)}$ is odd,
  then  $\gcd(e,{p}^{m}-1)\le \gcd (2({p}^{h}+1),{p}^{m}-1)\le 4$.
   Thus for any $0<j<m$, we have  $\gcd (e,{p}^{m}-1)\gcd ({p}^{j}-1,{p}^{m}-1)\le 4({p}^{\gcd(j,m)}-1)\le 4({p}^{\frac{m}{2}}-1)<{p}^{m}-1$
due to $p\ge 5$, thus  $\gcd (e,{p}^{m}-1)\gcd ({p}^{j}-1,{p}^{m}-1)\not\equiv 0\pmod{ {p}^{m}-1}$.
If $\frac{m}{\gcd (m,h)}$ is even and $h\ne \frac{m}{2}$, then $\gcd (m,h)\le \frac{m}{4}$.
Thus,
 $\gcd (e,{p}^{m}-1)\gcd ({p}^{j}-1,{p}^{m}-1)\le 2({p}^{\frac{m}{4}}+1)({p}^{\gcd(j,m)}-1)<({p}^{\frac{m}{4}}-1)({p}^{\frac{m}{4}}+1)({p}^{\frac{m}{2}}-1)={p}^{m}-1$, thus $|{C}_{e}|=m$.
%
\done
\begin{theorem}\label{ph+1thm}
	Let $p\ge 5$ be an odd prime and $m$ be a positive integer such that  ${p}^{m}\equiv 1\pmod 4$.
 Let  $s=\frac{{p}^{m}-1}{2}$ and  $e=\frac{{p}^{m}-1}{2}+{p}^{h}+1$.  Let    $h\ne\frac{m}{2}$ if $m$ is even,  then $\mathcal{C}_{(1,e,s)}$ is an  optimal $p$-ary cyclic code with parameters $[{p}^{m}-1,{p}^{m}-2m-2,4]$.
\end{theorem}
{\em Proof:}
According to    Lemma \ref{ph+1lem}, we have  $e\notin {C}_{1}$ and  $|{C}_{e}|=m$.
It is clearly that $\mathcal{C}_{(1,e,s)}$ does not have a  codeword of Hamming weight 1.
 In the following we show that
 $\mathcal{C}_{(1,e,s)}$ does not have a  codeword of Hamming weight 2 or 3.

   Suppose that
  $\mathcal{C}_{(1,e,s)}$ has a codeword of Hamming weight 2, then there exist two elements ${{c}_{1}}$ and
${{c}_{2}}\in {{\mathbb{F}}_{p}^*}$ and two distinct elements ${{x}_{1}} $ and
${{x}_{2}}\in {{\mathbb{F}}_{p^m}^*}$ such that (\ref{weight2equ}) is satisfied.
 If $({x}_{1}^{s},{x}_{2}^{s})=(1,1)$, we get  ${c}_{1}=-{c}_{2}$ and
  ${x}_{1}={x}_{2}$, which is contrary to ${x}_{1}\ne {x} _{2}$. If
  $({x}_{1}^{s},{x}_{2}^{s})=(1,-1)$, we have  ${c}_{1}={c}_{2}$
  and  ${x}_{1}=-{x}_{2}$.
  Then ${x}_{1}^{s}={(-{x}_{2})}^{s}={x}_{2}^{s} =1$ due to $\frac{p^m-1}{2}$ is even,  which is contrary to ${x}_{2}^{s}=-1$. Thus $\mathcal{C}_{(1,e,s)}$ does not have a  codeword of Hamming weight 2.

 The code $\mathcal{C}_{(1,e,s)}$ has  a codeword of  Hamming weight 3 if and only if   (\ref{weight3equ}) has no pairwise different nonzero  solutions ${x}_{1},{x}_{2},{x}_{3}$.
By the first two equations in (\ref{weight3equ}), we have
	\begin{equation}\label{ph+1eq1}
	{(-{c}_{2}{x}_{2}-{c}_{3}{x}_{3})}^{e}+{c}_{2}{x}_{2}^{e}+{c}_{3}{x}_{3}^{e}=0.
    \end{equation}
 Due  to symmetry,
 it is sufficient to consider the following two cases.

Case 1), 	  	$({x}_{1}^{s},{x}_{2}^{s},{x}_{3}^{s})=(1,1,1)$.
In this case, $1+{c}_{2}+{c}_{3}=0$ and $$({x}_{1}^{e},{x}_{2}^{e},{x}_{3}^{e})=({x}_{1}^{{p}^{h}+1},{x}_{2}^{{p}^{h}+1},{x}_{3}^{{p}^{h}+1}).$$
 Then (\ref{ph+1eq1}) becomes 	
		\begin{equation*}
			{(-{c}_{2}{x}_{2}-{c}_{3}{x}_{3})}^{{p}^{h}+1}+{c}_{2}{x}_{2}^{{p}^{h}+1}+{c}_{3}{x}_{3}^{{p}^{h}+1}=0, 
		\end{equation*}
which can be simplified to
		\begin{equation}\label{ph+1eq2}
({c}_{2}^{2}+{c}_{2}){x}_{2}^{{p}^{h}+1}+
({c}_{3}^{2}+{c}_{3}){x}_{3}^{{p}^{h}+1}+{c}_{2}{c}_{3}({x}_{2}{x}_{3}^{{p}^{h}}+{x}_{2}^{{p}^{h}}{x}_{3})=0.
		\end{equation}
By  ${c}_{2}+{c}_{3}=-1$, we have
 ${c}_{2}^{2}+{c}_{2}=-{c}_{2}{c}_{3}$ and ${c}_{3}^{2}+{c}_{3}=-{c}_{2}{c}_{3}$, then
 (\ref{ph+1eq2}) becomes
     ${x}_{2}^{{p}^{h}+1}+{x}_{3}^{{p}^{h}+1}-({x}_{2}{x}_{3}^{{p}^{h}}+{x}_{2}^{{p}^{h}}{x}_{3})=0$, i.e., ${({x}_{2}-{x}_{3})}^{{p}^{h}+1}=0$, which means  ${x}_{2}={x}_{3}$.
	
Case  2),
$({x}_{1}^{s},{x}_{2}^{s},{x}_{3}^{s})=(1,1,-1)$. In this case, $({x}_{1}^{e},{x}_{2}^{e},{x}_{3}^{e})=({x}_{1}^{{p}^{h}+1},{x}_{2}^{{p}^{h}+1},-{x}_{3}^{{p}^{h}+1})$, then  (\ref{ph+1eq1}) is equivalent to
	\begin{equation*}
		{(-{c}_{2}{x}_{2}-{c}_{3}{x}_{3})}^{{p}^{h}+1}+
{c}_{2}{x}_{2}^{{p}^{h}+1}-{c}_{3}{x}_{3}^{{p}^{h}+1}=0,
	\end{equation*}
which can be simplified to
	\begin{equation}\label{ph+1eq3}	({c}_{2}^{2}+{c}_{2}){x}_{2}^{{p}^{h}+1}+
({c}_{3}^{2}-{c}_{3}){x}_{3}^{{p}^{h}+1}+
{c}_{2}{c}_{3}({x}_{2}{x}_{3}^{{p}^{h}}+{x}_{2}^{{p}^{h}}{x}_{3})=0. 
	\end{equation}
By  $1+{c}_{2}-{c}_{3}=0$, we have  ${c}_{2}^{2}+{c}_{2}={c}_{2}{c}_{3}$ and
${c}_{3}^{2}-{c}_{3}={c}_{2}{c}_{3}$,
then (\ref{ph+1eq3}) becomes
${x}_{2}^{{p}^{h}+1}+{x}_{3}^{{p}^{h}+1}+({x}_{2}{x}_{3}^{{p}^{h}}+{x}_{2}^{{p}^{h}}{x}_{3})=0$, i.e.,  ${({x}_{2}+{x}_{3})}^{{p}^{h}+1}=0$, from which we  get  ${x}_{2}=-{x}_{3}$. Then  $1={x}_{2}^{s}={(-{x}_{3})}^{s}={x}_{3}^{s}$ due to
$s=\frac{p^m-1}{2}$ is even, which is contrary to  ${x}_{3}^{s}=-1$.


As a consequence,
   $\mathcal{C}_{(1,e,s)}$ does not have  a codeword of Hamming weight 3.
Then we complete the proof. 	
\done

\begin{remark}
	In 2020, Fan and Zhang \cite{fanzhagn20} constructed a class of  optimal quinary  cyclic code $\mathcal{C}_{(1,e,s)}$ with parameters  $[{5}^{m}-1,{5}^ {m}-2m-2,4]$, where  $e=\frac{5^m-1}{2}+5^h+1$ and $s=\frac{5^m-1}{2}$.  It is the
 special case $p=5$ of Theorem \ref{ph+1thm}.
\end{remark}

	\begin{example}
		Let $p=7$, $m=4$, and $h=3$. Then  $e=\frac{{7}^{4}-1}{2}+{7}^{3}+1=1544 $.
   Let $\alpha $ be the generator  of   $\mathbb{F}_{{7}^{4}}^*$
with
${\alpha}^{4}+5 {\alpha}^{2}+4\alpha+3=0$. Then the  code $\mathcal{C}_{(1,e,s)}$
 has  parameters $[2400,2391,4]$ and
 generator polynomial
   ${x}^{9}+5{x} ^{8}+6{x}^{7}+3{x}^{6}+3{x}^{4}+4{x}^{3}+2{x}^{2}+x+6$.
	\end{example}

	\begin{example}
		Let $p=11$, $m=2$, and  $h=0$. Then
 $e=\frac{{11}^{2}-1}{2}+{11}^{0}+		1=62$.
  Let $\alpha $ be the generator  of   $\mathbb{F}_{{11}^{2}}^*$
with
${\alpha}^{2}+7\alpha+2=0$. Then the  code $\mathcal{C}_{(1,e,s)}$
 has  parameters $[120,115,4]$ and
 generator polynomial
 ${x}^{5}+9{x}^{4}+10{x}^{3}+10{x}^{2}+5x+8$.
	\end{example}

\subsection{The third class of optimal $p$-ary cyclic  codes with minimal distance 4}

In this subsection, we will
 consider the exponents $e$ of the form
\begin{eqnarray}\label{econgre+}
  e(p^{k}+1) &\equiv& p^h+1\pmod{p^m-1},
\end{eqnarray}
where $0\leq  h,k\leq m-1$ are nonnegative integers.

\begin{lemma}\label{e1modp-1m}
  Let $p\ge 5$  be an odd prime and $m$ be a  positive
    integer.  Let $s=\frac{p^m-1}{2}$.
   Then the following system of equations has solutions
   $x,y\in\F_{p}\setminus\{0,1\}, x\ne y $ for some
     $c_1,c_2\in\F_{p}^*$:
   \begin{equation}\label{e1modp-1m1}
	\left\{ \begin{array}{l}
		1+c_1x+c_2y=0  \\
		1+{c}_{1}x^s+{c}_{2}y^s=0.
	\end{array} \right.
\end{equation}
\end{lemma}

{\em Proof:}
Since $x,y\in\F_{p}\setminus\{0,1\}$, we have $x^{p-1}=y^{p-1}=1$,
$x^{\frac{p-1}{2}}=\pm 1$, and $y^{\frac{p-1}{2}}=\pm 1$.
We consider the solution of (\ref{e1modp-1m1}) based on the parity of $m$.

Case 1),   $m$ is even.
In this case, $x^s=x^{(p-1)\frac{1+p+p^2+\cdots +p^{m-1}}{2}}=1$.
Similarly,  $ y^s=1$.
By (\ref{e1modp-1m1}),  $x=\frac{(1+c_1)y-1}{c_1}$. Let $c_1=1$ and $y=2$, then
$x=3$. Thus $(x,y)=(3,2)$ is a solution of  (\ref{e1modp-1m1}).

Case 2),   $m$ is odd.
In this case,    $x^s=x^{\frac{p-1}{2}{(1+p+p^2+\cdots +p^{m-1})}}=x^{\frac{p-1}{2}}$.
Similarly,  $ y^s=y^{\frac{p-1}{2}}$.
Let $( x^s,y^s)=(-1,-1)$, then from (\ref{e1modp-1m1}), we have
\begin{equation}\label{elmodp-1m2}
  y=\frac{1+c_1x}{c_1-1}.
\end{equation}
By (\ref{elmodp-1m2}),  $x=y$ will lead to $x=y=-1$.
Thus, (\ref{e1modp-1m1}) has  solutions
   $x,y\in\F_{p}\setminus\{0,1\} $ such that
   $x\ne y$ if
   \begin{equation}\label{e1modp-1m3}
	\left\{ \begin{array}{l}
		x^{\frac{p-1}{2}}=-1  \\
		(\frac{1+c_1x}{c_1-1})^{\frac{p-1}{2}}=-1
	\end{array} \right.
\end{equation}
has solutions in $x\in \F_{p}\setminus\{0,\pm 1\}$.
It can be seen that (\ref{e1modp-1m3})
has solutions in $x\in \F_{p}\setminus\{0,\pm 1\}$ is equivalent
to
   \begin{equation}\label{e1modp-1m4}
	\left\{ \begin{array}{l}
		x^{\frac{p-1}{2}}=-c_1^{\frac{p-1}{2}}  \\
		(x+1)^{\frac{p-1}{2}}=-(c_1-1)^{\frac{p-1}{2}}
	\end{array} \right.
\end{equation}
has solutions in $x\in \F_{p}\setminus\{0,\pm c_1\}$.

If $p\equiv 1\pmod 4$, we have $N_{-1,1}=\frac{p-1}{4}\ge 1$  by Lemma \ref{numofso}.
Thus, there exists  $c_1\in\F_{p}^*$ such that
$c_1^{\frac{p-1}{2}}=1$ and $(c_1-1)^{\frac{p-1}{2}}=-1$.  For this $c_1$,
(\ref{e1modp-1m4}) has  $N_{-1,1}=\frac{p-1}{4}$ solutions.
Note that $-1$ is a square in $\F_p^*$ for $p\equiv 1\pmod 4$. Thus,
$x=-c_1$
is not a solution of (\ref{e1modp-1m4}).
Clearly, $x=c_1$ is not a solution of (\ref{e1modp-1m4}).
As a result, (\ref{e1modp-1m4})
has a solution in  $x\in \F_{p}\setminus\{0,\pm c_1\}$  due to
 $ \frac{p-1}{4}\ge 1$.

If $p\equiv 3\pmod 4$, we have $N_{1,-1}=\frac{p+1}{4}\ge 2$  by Lemma \ref{numofso}.
Thus, there exists  $c_1\in\F_{p}^*$ such that
$c_1^{\frac{p-1}{2}}=-1$ and $(c_1-1)^{\frac{p-1}{2}}=1$.  For this $c_1$,
(\ref{e1modp-1m4}) has  $N_{1,-1}=\frac{p+1}{4}$ solutions.
For $p\equiv 3 \pmod 4$,  $-1$ is  a nonsquare in $\F_p^*$ . Thus,
$x=-c_1$
is  a solution of (\ref{e1modp-1m4}).
Clearly, $x=c_1$ is not a solution of (\ref{e1modp-1m4}).
As a result, (\ref{e1modp-1m4})
has a solution in  $x\in \F_{p}\setminus\{0,\pm c_1\}$  due to
 $ \frac{p+1}{4}\ge 2$.
This completes the proof.
  \done

\begin{lemma}\label{e1modp-1notoptimal}
  Let $p\ge 5$  be an odd prime and $m$ be a  positive
    integer.  Let $s=\frac{p^m-1}{2}$.
    If $e\equiv 1\pmod{p-1} $, then
    $\mathcal{C}_{(1,e,s)}$ has a codeword of weight 3.
\end{lemma}

{\em Proof:}
$\mathcal{C}_{(1,e,s)}$ has a codeword of weight 3 if and only if there exist $c_1,c_2\in\F_{p}^*$
and $x,y\in \F_{p^m}\setminus\{0,1\}$, $x\ne y$ such that

   \begin{equation}\label{e1modp-1notoptimal1}
	\left\{ \begin{array}{l}
		1+c_1x+c_2y=0  \\
           1+c_1x^e+c_2y^e=0  \\
		1+{c}_{1}x^s+{c}_{2}y^s=0.
	\end{array} \right.
\end{equation}
Now we restrict $x,y\in \F_{p}\setminus\{0,1\}$.
Since  $e\equiv 1\pmod{p-1}$, we have $x^e=x$ and  $y^e=y$.
By Lemma \ref{e1modp-1m}, (\ref{e1modp-1notoptimal1}) has a solution such that
$x,y\in \F_{p^m}\setminus\{0,1\}$ and  $x\ne y$.  \done

By Lemma \ref{e1modp-1notoptimal}, if
$\mathcal{C}_{(1,e,s)}$ has parameters $[p^m-1,p^m-2m-2,4]$,
then $ e\notequiv 1\pmod{p-1}$. Therefore, we will
restrict  $ e\notequiv 1\pmod{p-1}$ in order to construct optimal $p$-ary
cyclic codes with parameters $[p^m-1,p^m-2m-2,4]$.

\begin{theorem}\label{econgre+thm}
Let $p\ge 5$ be an odd prime and $m$ be a positive even integer.
Let $h,k$ be positive integers such that $\gcd(h-k,m)=1$ and
$\gcd(h+k,m)=1$.
  Let $e$ be  an integer defined by
  (\ref{econgre+}) such that
  $  e\notequiv 1\pmod {p-1}$.
  Let  $s=\frac{p^m-1}{2}$.
   Then $\mathcal{C}_{(1,e,s)}$ is
   an optimal $p$-ary cyclic code with
    parameters $[{p}^{m}-1,{p}^{m}-2m-2,4]$ if and only if  the following system of equations
    has no solution in $\F_p \setminus \{0,1\}$ for any $c_2,c_3\in\F_p^*$:

       \begin{equation}\label{econgre+eq1}
	\left\{ \begin{array}{l}
		x+c_2+c_3y=0  \\
        x^e+c_2+c_3y^e=0  \\
		1+{c}_{2}+{c}_{3}=0.
	\end{array} \right.
\end{equation}
\end{theorem}

{\em Proof:}
We first show that $e\notin C_1$ and $|C_e|=m$.
If $e\in C_1$, then $e\equiv p^i\pmod{p^m-1}$ for some $0 \le i\le m-1$.
Therefore  $e \equiv 1\pmod {p-1}$,
this is contrary to the assumption that $e \notequiv 1\pmod {p-1}$.    Then
$ e\notin C_1$.

 Note that
$  e(p^k+1)\equiv p^h+1\pmod{p^m-1}$.
If $\frac{m}{\gcd(m,h)}$ is odd, then $\gcd(e,p^m-1)\le \gcd(e(p^k+1),p^m-1)=
\gcd(p^h+1,p^m-1)=2.$ According to Lemma \ref{ce}, $|C_e|=m$.
If $\frac{m}{\gcd(m,h)}$ is even, then $\gcd(e,p^m-1)\le  \gcd(e(p^k+1),p^m-1)=
\gcd(p^h+1,p^m-1)=p^{\gcd(m,h)}+1.$
Suppose that $|C_e|=l_e$. Then $e(p^{l_e}-1)\equiv 0\pmod{p^m-1}$.
Thus,
$p^m-1|\gcd(e,p^m-1)\gcd(p^{l_e}-1,p^m-1)$.
If $l_e\ne m$, then $l_e\le \frac{m}{2}$.
Thus,
$p^m-1\le \gcd(e,p^m-1)\gcd(p^{l_e}-1,p^m-1)\le (p^{\gcd(m,h)}+1)(p^{\gcd(l_e,m)}-1 )
\le (p^{\frac{m}{2}}+1)(p^{\frac{m}{2}}-1)=p^m-1$,
which lead to $\gcd(e,p^m-1)=p^{\gcd(m,h)}+1=p^{\frac{m}{2}}+1 $ and $l_e=\frac{m}{2}$.
By
$\gcd(e(p^k+1),p^m-1)=p^{\gcd(m,h)}+1
=p^{\frac{m}{2}}+1,$  we have
 $\gcd(\frac{e}{p^\frac{m}{2}+1}(p^k+1),p^{\frac{m}{2}}-1)=
1,$ which is impossible.  Thus $\l_e=m$.

  Suppose that
  $\mathcal{C}_{(1,e,s)}$ has a codeword of Hamming weight 2, then there exist two elements ${{c}_{1}}$ and
${{c}_{2}}\in {{\mathbb{F}}_{p}^*}$ and two distinct elements ${{x}_{1}} $ and
${{x}_{2}}\in {{\mathbb{F}}_{p^m}^*}$ such that (\ref{weight2equ}) is satisfied.
 If $({x}_{1}^{s},{x}_{2}^{s})=(1,1)$, we get  ${c}_{1}=-{c}_{2}$ and
  ${x}_{1}={x}_{2}$, which is contrary to ${x}_{1}\ne {x} _{2}$. If
  $({x}_{1}^{s},{x}_{2}^{s})=(1,-1)$, we have  ${c}_{1}={c}_{2}$
  and  ${x}_{1}=-{x}_{2}$.
  Then $1={x}_{1}^{s}={(-{x}_{2})}^{s}={x}_{2}^{s} $ since  $p^m\equiv 1\pmod 4$ for even $m$,   which is contrary to ${x}_{2}^{s}=-1$. Thus $\mathcal{C}_{(1,e,s)}$ does not have a  codeword of Hamming weight 2.

Suppose that
  $\mathcal{C}_{(1,e,s)}$ has a codeword of Hamming weight 3, then there exist two elements ${{c}_{2}}$ and
${{c}_{3}}\in {{\mathbb{F}}_{p}^*}$ and three distinct elements ${{x}_{1}} $,  ${{x}_{2}} $, and
${{x}_{3}}\in {{\mathbb{F}}_{p^m}^*}$ such that (\ref{weight3equ}) is satisfied.
Let $x=\frac{x_1}{x_2}$ and $y=\frac{x_3}{x_2}$,
then $x,y\notin\{ 0,1\}$,  $x\ne y$, and
(\ref{weight3equ}) can be rewritten as
          \begin{equation}\label{econgre+eq2}
	\left\{ \begin{array}{l}
		x+c_2+c_3y=0  \\
        x^e+c_2+c_3y^e=0  \\
		x^s+{c}_{2}+{c}_{3}y^s=0.
	\end{array} \right.
\end{equation}
From the first two equations in (\ref{econgre+eq2}),
we have
\begin{equation}\label{econgre+eq3}
   (-c_2-c_3y)^e=-c_2-c_3y^e.
\end{equation}
 Taking $(p^k+1)$-th power on both sides of (\ref{econgre+eq3}) will  lead  to
$ (c_2+c_3y)^{e(p^k+1)}=(c_2+c_3y^e)^{p^k+1}$. By $e(p^k+1)\equiv p^h+1\pmod{p^m-1}$, we have
$ (c_2+c_3y)^{p^h+1}=(c_2+c_3y^e)^{p^k+1}$, which is equivalent to
$y^{p^h}+y=y^{e\cdot p^k}+y^e$. Note that
$ e\cdot p^k \equiv p^h+1-e \pmod{p^m-1}$, we have
$y^{p^h}+y=y^{p^h+1-e }+y^e$, i.e.,
$y(y^{p^h-e}-1)(y^{e-1}-1)=0$.
Therefore, $ y^{p^h-e}=1$ or  $y^{e-1}=1$. Note that
$\gcd(p^h-e,p^m-1)| \gcd((p^h-e)(p^k+1),p^m-1)=\gcd(p^{h+k}-1,p^m-1)=
p^{\gcd(h+k,m)}-1=p-1$ by $\gcd(h+k,m)=1$, and
$\gcd(e-1,p^m-1)| \gcd((e-1)(p^k+1),p^m-1)=\gcd(p^{h}-p^k,p^m-1)=
p^{\gcd(h-k,m)}-1=p-1$ by $\gcd(h-k,m)=1$.
Therefore we have $y^{p-1}=1$,  i.e., $y\in \F_{p}^* $ and thus
$y^s=y^{\frac{p^m-1}{2}}=y^{(p-1)\frac{1+p+\cdots +p^{m-1}}{2}}=1$.
Similarly, we have  $x=-c_2-c_3y\in\F_p^*$ and $x^s=1$.
Thus, (\ref{econgre+eq2}) becomes (\ref{econgre+eq1}).
Therefore, $\mathcal{C}_{(1,e,s)}$ has no codeword of weight 3 if and only if
 (\ref{econgre+eq1})
has no solution in $\F_p \setminus \{0,1\}$ for any $c_2,c_3\in\F_p^*$.
This completes the proof.
\done

Similarly, let $e$ be a solution of
\begin{eqnarray}\label{econgre-}
  e(p^{k}-1) &\equiv& p^h-1\pmod{p^m-1}
\end{eqnarray}
such that $e\notequiv 1\pmod {p-1}$, we have the
following theorem.

\begin{theorem}\label{econgre-thm}
Let $p\ge 5$ be an odd prime and $m$ be a positive even integer.
Let $h,k$ be positive integers such that $\gcd(h,m)=1$ and
$\gcd(h-k,m)=1$.
  Let $e$ be  an integer defined by
  (\ref{econgre-}) such that
  $  e\notequiv 1\pmod {p-1}$.
  Let  $s=\frac{p^m-1}{2}$.
   Then $\mathcal{C}_{(1,e,s)}$ is
   an optimal $p$-ary cyclic code with
    parameters $[{p}^{m}-1,{p}^{m}-2m-2,4]$ if and only if   (\ref{econgre+eq1})
    has no solution in $\F_p \setminus \{0,1\}$ for any $c_2,c_3\in\F_p^*$:
\end{theorem}

\begin{remark}
  It is usually easy to determine the solutions 
  of (\ref{econgre+eq1}) due to  $x,y\in\F_{p}^*$. We give an example in the case  $p=5$ as follows.
\end{remark}

\begin{theorem}\label{p5h+1div2thm}
Let $p=5$ and $m$ be an even integer.  Let  $e=\frac{{5}^{m}-1}{2}+\frac{5^h+1}{2}$,
   where $\gcd(h,m)=1 $.    Then  $\mathcal{C}_{(1,e,s)}$ has parameters  $[{5}^{m}-1,{5}^{m}-2m-2,4]$.
\end{theorem}
{\em Proof:}
Note that $e(5^0+1)\equiv 5^h+1\pmod{5^m-1}$, thus $e$ is a solution of $(\ref{econgre+})$.
  Since $m$ is even and $h$   is odd due to  $\gcd(h,m)=1$,  we have $\frac{5^m-1}{2}\equiv 0\pmod 4$ and
  $\frac{5^h+1}{2}\equiv 3\pmod 4$. Thus,   $e\equiv 3\pmod 4$.
    According to Theorem \ref{econgre+thm}, we only need to show that
  (\ref{econgre+eq1})
   has no solution in $\F_p \setminus \{0,1\}$ for any $c_2,c_3\in\F_p^*$.
  Notice that $y\in \F_p \setminus \{0,1\}=\{-1,\pm2\}$.

   If $y=-1$,  then $y^e=-1$, and
 thus $x+c_2-c_3=x^e+c_2-c_3=0$, i.e., $x=x^e$. Together with $x^{p-1}=x^4=1$,
  we have $x^{\gcd(e-1,p-1)}=x^2=1$ by $e\equiv 3\pmod 4$, i.e., $x=\pm1$. Then $x=-1$ since  $x \in \F_p \setminus \{0,1\}$.
 As a consequence,   the first equation of   (\ref{econgre+eq1})  becomes $-1+c_2-c_3=0$,  together with the third equation of
  (\ref{econgre+eq1}),  we have $c_2=0$.

  If $y=\pm2$, then $y^2=-1$. Since $e\equiv 3\pmod4$,
    (\ref{econgre+eq1}) can be reduced to
             \begin{equation}\label{p5h+1div2eq1}
	\left\{ \begin{array}{l}
		x+c_2+c_3y=0  \\
        x^3+c_2-c_3y=0  \\
		1+{c}_{2}+{c}_{3}=0.
	\end{array} \right.
\end{equation}
    Substituting  $x=-c_2-c_3y  $ into $x^3+c_2-c_3y=0$
  yields $(c_2+c_3y)^3+c_3y-c_2=0$, which can be reduced to
  $(c_2-1)y=c_3-1$. Note that $1+c_2+c_3=0$, we have $y=-\frac{c_2+2}{c_2-1}=\pm2$,
  i.e., $c_2=0$ or $c_2=-1$. However $c_2=-1$ means $ c_3=0$.

  To sum up,  (\ref{econgre+eq1}) has no
   solution in $\F_p \setminus \{0,1\}$ for any $c_2,c_3\in\F_p^*$.
    \done

\begin{example}
Let $p=5$,      $m=4$, and $h=1$.  Then  $e=\frac{5^4-1}{2}+3=315$.
   Let $\alpha $ be the generator  of   $\mathbb{F}_{{5}^{4}}^*$
with
$\alpha^4 + 4\alpha^2 + 4\alpha + 2=0$. Then the  code $\mathcal{C}_{(1,e,s)}$
 has  parameters $[624,615,4]$ and
 generator polynomial
 $x^9 + 4x^8 + 2x^5 + 2x^3 + 2x^2 + 2x + 1$.
\end{example}

\begin{example}
Let $p=5$,    $m=4$, and $h=3$.  Then  $e=\frac{5^4-1}{2}+\frac{5^3+1}{2}=375$.
   Let $\alpha $ be the generator  of   $\mathbb{F}_{{5}^{4}}^*$
with
$\alpha^4 + 4\alpha^2 + 4\alpha + 2=0$. Then the  code $\mathcal{C}_{(1,e,s)}$
 has  parameters $[624,615,4]$ and
 generator polynomial
 $x^9 + 3x^8 + 4x^7 + 2x^6 + 3x^4 + 3x^3 + 3x^2 + 4x + 1$.
\end{example}

%
%
%

\subsection{The fourth  class of optimal $p$-ary cyclic  codes with minimal distance 4}

In this subsection, we will consider the exponent $e$ of the form
 $e=\frac{{p}^{m}-1}{2}-1$.

\begin{lemma}\label{pm-1lem}
 Let $p\ge 5$ be an odd prime and $m$ be a positive integer.
 Let $e=\frac{{p}^{m}-1}{2}-1$. Then
 $e\notin {C}_{1}$ and  $|{C}_{e}|=m$.
\end{lemma}
{\em Proof:}
  Since $\gcd (\frac{{p}^{m}-1}{2}-1,{p}^{m}-1)\leq 2$ by  $\gcd(2(\frac{{p}^{m}-1}{2}-1),{p}^{m}-1)=2$.
  Thus  $|{C}_{e}|=m$ by Lemma \ref{ce}.
Suppose that   $e\in {C}_{1}$,  then there exists a positive integer $i$
  such that
$\frac{{p}^{m}-1}{2}-1\equiv {p}^{i}\pmod{{p}^{m}-1}$, then ${p}^{m}-1|\frac{{p}^{m}-1}{2}-1-{p}^{i}$, we get ${p}^{m}-1|2(1+{p}^{i})$.
Remember that
 $$\gcd({p}^{i}+1,{p}^{m}-1)=\left\{
\begin{array}{ll}
2,& \textup{if}\ \frac{m}{\gcd(m,i)}\  \textup{is odd},\\
p^{\gcd(m,i)}+1,& \textup{if}\ \frac{m}{\gcd(m,i)}\  \textup{is even}.
\end{array}
\right.$$
 Thus $\gcd ({p}^{i}+1,{p}^{m}-1)\leq {p}^{\frac{m}{2}}+1$. By ${p}^{m}-1|2(1+{p}^{i})$ and
 ${p}^{m}-1|2({p}^{m}-1)$, we have ${p}^{m}-1|2\gcd(p^i+1,p^m-1)$, then ${p}^{m}-1\le 2({p}^{\frac{m}{2}}+1)$,
 i.e.,
  ${p}^{\frac{m}{2}}-1\leq 2$, which is impossible for $p\ge 5$, thus $e\notin {C}_{1}$.
\done

 \begin{theorem}\label{pm-1thm}
 	Let $p\ge 5$ be an odd prime and $m$ be a positive integer.
  Let  $e=\frac{{p}^{m}-1}{2}-1$ and $s=\frac{p^m-1}{2}$.
  If for each $a\in \mathbb{F}_{p}\backslash \{0,-2\}$,  one of the following three conditions is satisfied,
 	then  the code  $\mathcal{C}_{(1,e,s)}$ is an optimal $p$-ary cyclic code with parameters $[{p}^{m}-1,{p}^{m}-2m-2,4]$:
 	\begin{itemize}
 		\item [1)]
 		   ${a}^{2}+4$ is a nonsquare of $\mathbb{F}_{{p}^{m}}$; or
 		\item [2)]
 		    $\eta (\frac{-a\pm \sqrt{{a}^{2}+4}}{2})\ne -1$; or
 		\item [3)]
 		$\eta (-\frac{2}{a}\cdot \frac{-a\pm \sqrt{{a}^{2}+4}}{2}-\frac{2}{a}-1)\ne -1$.
 	\end{itemize}
 \end{theorem}
{\em Proof:}
  According to Lemma \ref{pm-1lem}, we have  $| \mathcal{C}_{e} |=m$ and $e\notin {C}_{1}$. Clearly, the minimal
   distance $d$    of the code  $\mathcal{C}_{(1,e,s)}$  cannot be  1.

   Suppose that  $\mathcal{C}_{(1,e,s)}$ has  a  codeword of Hamming weight 2.
   Then  there exist two elements ${{c}_{1}}$ and
${{c}_{2}}\in {{\mathbb{F}}_{p}^*}$ and two distinct elements ${{x}_{1}} $ and
${{x}_{2}}\in {{\mathbb{F}}_{p^m}^*}$ such that
(\ref{weight2equ}) holds.
  If $({x}_{1}^{s},{x}_{2}^{s})=(1,1)$,
  we have ${c}_{1}=-{c}_{2}$ and ${x}_{1}={x}_{2}$, which is contrary to ${x}_{1}\ne {x}_{2}$.
  If $({x}_{1}^{s},{x}_{2}^{s})=(1,-1)$,  by the   third equation of   (\ref{weight2equ}),
   we have  ${c}_{1}={c}_{2}$. 
   From the second equation of (\ref{weight2equ}), we have
   $c_1x_1^e+c_2x_2^e=c_1(x_1^{-1}-x_2^{-1})=0$, i.e., $x_1=x_2$.
 Thus $\mathcal{C}_{(1,e,s)}$ does not have  a codeword of Hamming weight 2.

    Next we show that    $\mathcal{C}_{(1,e,s)}$ has no codeword of weight 3.
   Due to symmetry, it is sufficient to consider the following two cases.

Case  1),     $({x}_{1}^{s},{x}_{2}^{s},{x}_{3}^{s})=(1,1,1)$.
In this case, (\ref{weight3equ}) becomes
\begin{equation}\label{pm-1eq1}
	\left\{ \begin{array}{l}
		{x}_{1}+{c}_{2}{x}_{2}+{c}_{3}{x}_{3}=0  \\
		{x}_{1}^{-1}+{c}_{2}{x}_{2}^{-1}+{c}_{3}{x}_{3}^{-1}=0  \\
		1+{c}_{2}+{c}_{3}=0.
	\end{array} \right.
\end{equation}
From the first equation of (\ref{pm-1eq1}), we get ${x}_{1}=-({c}_{2}{x}_{2}+{c}_{3}{x}_{3})$, substitute it
into the second equation of (\ref{pm-1eq1}), we get
\begin{equation}\label{pm-1eq2}
	{-({c}_{2}{x}_{2}+{c}_{3}{x}_{3})}^{-1}+{c}_{2}{x}_{2}^{-1}+{c}_{3}{x}_{3}^{-1}=0.
\end{equation}
Multiplying both sides of (\ref{pm-1eq2})  by $({c}_{2}{x}_{2}+{c}_{3}{x}_{3})$ will lead to
\begin{equation*}
	-1+({c}_{2}{x}_{2}^{-1}+{c}_{3}{x}_{3}^{-1})({c}_{2}{x}_{2}+{c}_{3}{x}_{3})=0,
\end{equation*}
which can be simplified to
$
{c}_{2}{c}_{3}({x}_{3}^{-1}{x}_{2}+{x}_{2}^{-1}{x}_{3})+{c}_{2}^{2}+{c}_{3}^{2}-1=0.
$
Let ${x}_{2}^{-1}{x}_{3}=y$, then $y\notin \{0,1\}.  $ We have
\begin{equation}\label{pm-1eq3}
{y}^{2}+\frac{{c}_{2}^{2}+{c}_{3}^{2}-1}{{c}_{2}{c}_{3}}y+1=0.
\end{equation}
By  $1+{c}_{2}+{c}_{3}=0$,  we get $\frac{{c}_{2}^{2}+{c}_{3}^{2}-1}{{c}_{2}{c}_{3}}=
\frac{{c}_{2}^{2}+{(1+{c}_{2})}^{2}-1}{{c}_{2}{c}_{3}}=\frac{2{c}_{2}^{2}+2{c}_{2}}{-{c}_{2}(1+{c}_{2})}=-2$.
thus, (\ref{pm-1eq3}) becomes  ${y}^{2}-2y+1=0$, i.e.,  $y=1$,  which is contrary to $y\notin \{0,1\}.  $
	
Case 2),
   $({x}_{1}^{s},{x}_{2}^{s},{x}_{3}^{s})=(1,1,-1)$.
In this case, (\ref{weight3equ}) becomes
\begin{equation}\label{pm-1eq4}
	\left\{ \begin{array}{l}
		{x}_{1}+{c}_{2}{x}_{2}+{c}_{3}{x}_{3}=0  \\
		{x}_{1}^{-1}+{c}_{2}{x}_{2}^{-1}-{c}_{3}{x}_{3}^{-1}=0  \\
		1+{c}_{2}-{c}_{3}=0,
	\end{array} \right.
\end{equation}
By the first two equations
of (\ref{pm-1eq4}), we get
\begin{equation*}
	-{({c}_{2}{x}_{2}+{c}_{3}{x}_{3})}^{-1}+{c}_{2}{x}_{2}^{-1}-{c}_{3}{x}_{3}^{-1}=0.
\end{equation*}
Multiplying both sides of the above equation  by $({c}_{2}{x}_{2}+{c}_{3}{x}_{3})$ will lead to 
\begin{equation*}
	-1+({c}_{2}{x}_{2}^{-1}-{c}_{3}{x}_{3}^{-1})({c}_{2}{x}_{2}+{c}_{3}{x}_{3})=0,
\end{equation*}
i.e.,
\begin{equation}\label{pm-1eq5}
	{c}_{2}{c}_{3}({x}_{2}^{-1}{x}_{3}-{x}_{3}^{-1}{x}_{2})+{c}_{2}^{2}-{c}_{3}^{2}-1=0.
\end{equation}
Let ${x}_{3}^{-1}{x}_{2}=y$,
 then $\frac{x_1}{x_3}=\frac{-c_2x_2-(1+c_2)x_3}{x_3}=-(c_2y+c_2+1)$.
 Thus (\ref{pm-1eq5}) can be rewritten
  as
\begin{equation*}
	{y}^{2}-\frac{{c}_{2}^{2}-{c}_{3}^{2}-1}{{c}_{2}{c}_{3}}y-1=0.
\end{equation*}
By $1+{c}_{2}-{c}_{3}=0$,
 we get $\frac{{c}_{2}^{2}-{c}_{3}^{2}-1}{{c}_{2}{c}_{3}}=
 \frac{{c}_{2}^{2}-{(1+{c}_{2})}^{2}-1}{{c}_{2}{c}_{3}}=
 \frac{-2{c}_{2}-2}{{c}_{2}(1+{c}_{2})}=-\frac{2}{{c}_{2}}$.
As a result, we have
$
	{y}^{2}+\frac{2}{{c}_{2}}y-1=0,
$
where $c_2\notin \{0,-1\}$.
Let $\frac{2}{{c}_{2}}=a$, then  $a=\frac{2}{{c}_{2}}\in \mathbb{F}_{p}\backslash \{0,-2\}$.
 Note that  $ \eta(y)=\eta(\frac{x_2}{x_3})=-1 $ and  $ \eta(\frac{x_1}{x_3})=
 \eta(-(c_2y+c_2+1))=-1$.
 It is known that    the solutions of ${y}^{2}+ay-1=0$ are ${y_{1,2}}=\frac{-a\pm\sqrt{{a}^{2}+4}}{2}$.
Thus, (\ref{pm-1eq4}) has  solutions  if and only if  for some $a\in \mathbb{F}_{p}\backslash \{0,-2\}$, the following three conditions    are all satisfied:
\begin{itemize}
	\item [1)]
	${a}^{2}+4$ is a square of $\mathbb{F}_{{p}^{m}}$;
	\item [2)]
	$\eta ({y}_{1,2})=-1$; and
	\item [3)]
	$\eta (-\frac{2}{a}{y}_{1,2}-\frac{2}{a}-1)= -1$.
\end{itemize}

Therefore, if
 for each $a\in \mathbb{F}_{p}\backslash \{0,-2\}$,  one of the  three conditions in Theorem \ref{pm-1thm} is satisfied,
 $\mathcal{C}_{(1,e,s)}$ does not have a codeword of Hamming weight 3.
 This completes the proof. \done

 \begin{corollary}
	Let $p=5$ and $m$  be even. Let $e=\frac{{p}^{m}-1}{2}-1$, then $\mathcal{C}_{(1,e,s)}$ is an optimal cyclic code with parameters $[{p}^{m}-1,{p}^{m}-2m-2,4]$.
\end{corollary}
{\em Proof:}
By Theorem \ref{pm-1thm}, we need to  show that
 for each $a\in \mathbb{F}_{5}\backslash \{0,-2\}=\{\pm1,2\}$,  one of the  three conditions given in Theorem \ref{pm-1thm} is satisfied.

	Case 1),   $a= \pm 1$. In this case,  $\eta(\frac{-a \pm \sqrt{a^2+4}}{2})=\eta(\frac{\mp 1\pm \sqrt{5}}{2})=\eta (\pm 2)={(\pm 2)}^{\frac{{5}^{m}-1}{2}}=1$ due to $m$ is even.

 Case 2),   $a=2$. Then  $\frac{-a\pm \sqrt{a^2+4}}{2}=\frac{-2\pm 2\sqrt{2}}{2}=-1\pm \sqrt{2}$. It can be checked that the order of $-1\pm \sqrt{2}$ is 12.
Then $\eta (-1\pm \sqrt{2})={(-1\pm \sqrt{2})}^{\frac{{5}^{m}-1}{2}}=1$ by $12|\frac{{5}^{m}-1}{2}$ due to $m$ is even.
\done

\begin{example}
	Let $p=5$ and  $m=4$.  Then  $e=\frac{{5}^{4}-1}{2}-1=311$.
Let $\alpha $ be the generator  of   $\mathbb{F}_{{5}^{4}}^*$
with
${\alpha}^{4}+4{\alpha}^{2}+4\alpha+2=0$. Then  $\mathcal{C}_{(1,e,s)}$
 has  parameters $[624,615,4]$ and
 generator polynomial
${x}^{9}+4{x}^{8}+4{x}^{7}+2{x}^{6}+{x}^{5}+4{x}^{4}+3{x}^{2}+3x+1$.
\end{example}

\begin{corollary}
Let $p=7$ and $m$ be an odd integer or  $m\equiv 0\pmod 4$.
 Let $e=\frac{{p}^{m}-1}{2}-1$ and $s=\frac{p^m-1}{2}$. Then $\mathcal{C}_{(1,e,s)}$ is an  optimal cyclic code with parameters $[{p}^{m}-1,{p}^{m}-2m-2,4]$.
\end{corollary}	

{\em Proof:}
By Theorem \ref{pm-1thm}, we  shall  show that
 for each $a\in \mathbb{F}_{7}\backslash \{0,-2\}=\{\pm1,2,\pm 3\}$,  one of the  three conditions given in Theorem \ref{pm-1thm} is satisfied.

Case 1),   $a=\pm 1$.
  In this case, $\frac{-a\pm \sqrt{a^2+4}}{2}=\frac{\mp 1\pm \sqrt{5}}{2}$.
   If $m$ is odd, then $5$ is a nonsquare in $\mathbb{F}_{{7}^{m}}$.
   If $4|m$, then  $\eta (\frac{\mp 1\pm \sqrt{5}}{2})={(\frac{\mp 1\pm \sqrt{5}}{2})}^{\frac{{7}^{m}-1}{2}}=1$ due to the order of $\frac{\mp 1\pm \sqrt{5}}{2}$ is 48  and $48|\frac{{7}^{m}-1}{2}$.

Case 2),   $a=2$. We have  $\frac{-a\pm \sqrt{a^2+4}}{2}=\frac{-2\pm 2\sqrt{2}}{2}=-1\pm \sqrt{2}=-1\pm 4=2$ or 3.
If $m$ is   odd, then  $\eta (3)={3}^{\frac{{7}^{m}-1}{2}}=-1$, but   $\eta (-\frac{2}{a}\cdot 3-\frac{2}{a}-1)=\eta (2)={2}^{\frac{{7}^{m}-1}{2}}=1.$  If  $4|m$, then $\eta (2)=\eta (3)=1$.

Case 3),    $a=\pm 3$.
 In this case,  $\frac{-a\pm \sqrt{a^2+4}}{2}=\frac{\mp3\pm\sqrt{-1}}{2}$.  If
   $m$ is odd, then $-1$ is  a nonsquare of $\mathbb{F}_{{7}^{m}}$. If $4|m$, it can be checked that the order of $\frac{\mp3\pm\sqrt{-1}}{2}$ is 16.
Thus $\eta (\frac{\mp3\pm\sqrt{-1}}{2})=1$ by $16|\frac{{7}^{m}-1}{2}$ due to $4|m$.
\done

\begin{example}
	Let $p=7$ and  $m=3$. Then  $e=\frac{{7}^{3}-1}{2}-1=170$.
 Let $\alpha $ be the generator  of   $\mathbb{F}_{{7}^{3}}^*$
with
${\alpha}^{3}+6{\alpha}^{2}+4=0$. Then  $\mathcal{C}_{(1,e,s)}$
 has  parameters $[342,335,4]$ and
 generator polynomial
${x}^{7}+4{x}^{5}+2{x}^{4}+6{x}^{3}+{x}^{2}+5x+6$.
 \end{example}

\section{Two   classes of  optimal quinary  cyclic codes}

In this section, we will give two  classes of optimal quinary  cyclic codes
with parameters $[5^m-1,5^m-2m-2,4]$.

\begin{proposition}\cite{fanzhagn20}\cite{liucao20}\label{p5thm}
	Let $p=5$ and $m$ be a positive integer. Let $s=\frac{5^m-1}{2}$.
 Suppose that $e\notin {C}_{1}$ and $|{C}_{e}|=m$.
 Then $\mathcal{C}_{(1,e,s)}$ has parameters $[{5}^{m}-1,{5}^{m}-2m-2,4]$
 if and only if the following three systems of equations  have no solution in $\mathbb{F}_{{5}^{m}}\backslash \{0,1\}$:
	\begin{equation}\label{p51}
	\left\{
\begin{array}{l}
\eta(x)=-1\\
\eta(2x-2)=-1\\
(2(1-x))^e+2x^e-2=0,
\end{array}
\right.
	\end{equation}

		\begin{equation}\label{p52}
	\left\{
\begin{array}{l}
\eta(x)=1\\
\eta(2x+2)=1\\
(-2(1+x))^e+2x^e+2=0,
\end{array}
\right.
	\end{equation}
and
	\begin{equation}\label{p53}
	\left\{
\begin{array}{l}
\eta(x)=1\\
\eta(2x+2)=-1\\
(2(1+x))^e-2x^e-2=0.
\end{array}
\right.
	\end{equation}
\end{proposition}

\begin{theorem}\label{p5e3cor}
	Let $p=5$, $e\equiv 3\pmod 4$, and  $|{C}_{e}|=m$. Then $\mathcal{C}_{(1,e,s)}$
has parameters  $[{5}^{m}-1,{5}^{m}-2m-2,4]$  if and only if
 $ {(1+x)}^{e}+{x}^{e}+1=0$ has
 no solution in $\mathbb{F}_{{5}^{m}}\backslash \mathbb{F}_{5}$.
\end{theorem}

{\em Proof:}
Since $e\equiv 3\pmod 4$, thus $e\notin {C}_{1}$.
In fact, if  $e\in {C}_{1}$, then $ep^i\equiv 1 \pmod{p-1}$,
which implies that
$e\equiv 1 \pmod 4$.
Thus, by Proposition \ref{p5thm}, we need to show that
(\ref{p51})-(\ref{p53}) have no solution in $\F_{5^m}\backslash \{0,1\} $ if and only if
 $ {(1+x)}^{e}+{x}^{e}+1=0$ has no  solution in $\mathbb{F}_{{5}^{m}}\backslash \mathbb{F}_{5}$.

Since $e\equiv 3 \pmod 4$, we have
 ${2}^{e}\equiv 3\pmod 5$, then  (\ref{p51}) is equivalent to
\begin{equation}\label{p5e31a}
	\left\{ \begin{array}{l}
		\eta (x)=-1\\
\eta (2(x-1))=-1  \\
		{(x-1)}^{e}+{x}^{e}-1=0,
	\end{array} \right.
\end{equation}
where $x\notin \{ 0,1\}$. Note that $\eta(-1)=1$ and $\eta(2(-2-1))=1$, so that $x\notin \{ -1,-2\}$, and
$x=2$ is not a solution of ${(x-1)}^{e}+{x}^{e}-1=0$. Thus, $x\notin \F_5$.

(\ref{p52})
 is equivalent to
\begin{equation}\label{p5e32a}
\left\{
	 \begin{array}{l}
	\eta (x)=1\\
\eta (2(x+1))=1  \\
	{(1+x)}^{e}+{x}^{e}+1=0,
\end{array} \right.
\end{equation}
where $x\notin \{ 0,1,-1\}$. Note that
$x=\pm 2$ are not  solutions of ${(1+x)}^{e}+{x}^{e}+1=0$. Thus, $x\notin \F_5$.

(\ref{p53})
 is equivalent to
\begin{equation}\label{p5e32b}
	\left\{ \begin{array}{l}
	\eta (x)=1  \\
\eta (2(x+1))=-1  \\
	{(1+x)}^{e}+{x}^{e}+1=0,
\end{array} \right.
\end{equation}
where $x\notin \{ 0,1,-1\}$. Note that
$x=\pm 2$ are not  solutions of ${(1+x)}^{e}+{x}^{e}+1=0$. Thus, $x\notin \F_5$.

It can be seen that both (\ref{p5e32a}) and (\ref{p5e32b})
have no solutions in  $\mathbb{F}_{{5}^{m}}\backslash \mathbb{F}_{5}$ if and only if
\begin{equation}\label{p5e31}
\left\{ \begin{array}{l}
		\eta (x)=1  \\
		{(x+1)}^{e}+{x}^{e}+1=0
	\end{array} \right.
\end{equation}
has no solutions in  $\mathbb{F}_{{5}^{m}}\backslash \mathbb{F}_{5}$.
In the following, we show that (\ref{p5e31a}) has a  solution in $\mathbb{F}_{{5}^{m}}\backslash \mathbb{F}_{5}$ if and only if the following system of equations
 has a solution  in $\mathbb{F}_{{5}^{m}}\backslash \mathbb{F}_{5}$:
\begin{equation}\label{p5e31b}
	\left\{ \begin{array}{l}
	\eta (x)=-1\\
\eta (2(x-1))=1  \\
	{(x-1)}^{e}+{x}^{e}-1=0.
\end{array} \right.
\end{equation}
Actually, if  ${x}_{0}\in \mathbb{F}_{{5}^{m}}\backslash \mathbb{F}_{5}$
  is a solution of (\ref{p5e31a}), then
  $\frac{1}{{x}_{0}}\in \mathbb{F}_{{5}^{m}}\backslash \mathbb{F}_{5}$ is a solution of (\ref{p5e31b})
  due to
  $\eta (2(\frac{1}{{x}_{0}}-1))=-\eta (2({x}_{0}-1))$  and
  $\eta (\frac{1}{{x}_{0}})=\eta ({x}_{0})$.
Thus,  (\ref{p5e31a}) has no solution in $\mathbb{F}_{{5}^{m}}\backslash \mathbb{F}_{5}$
is  equivalent to
 \begin{equation}\label{p5e32}
\left\{ \begin{array}{l}
		\eta (x)=-1  \\
		{(x-1)}^{e}+{x}^{e}-1=0.
	\end{array} \right.
 \end{equation}
  has no solution in $\mathbb{F}_{{5}^{m}}\backslash \mathbb{F}_{5}$.
 Let $y=-x$ in $ (\ref{p5e32})$, then ${(x-1)}^{e}+{x}^{e}-1=0$ becomes
  ${(y+1)}^{e}+{y}^{e}+1=0$. Since $\eta(y)=\eta(-x)=\eta(x)$ by $\eta(-1)=1$, then
  $(\ref{p5e32})$  has no solution in $\mathbb{F}_{{5}^{m}}\backslash \mathbb{F}_{5}$ if and only if
  ${(y+1)}^{e}+{y}^{e}+1=0$ has no solution in $\mathbb{F}_{{5}^{m}}\backslash \mathbb{F}_{5}$
such that $ \eta(y)=-1$.
Together with (\ref{p5e31}), we get the conclusion.
\done

\begin{remark}
  For $p=5$ and $e\equiv 3\pmod 4 $, Theorem \ref{p5e3cor} gives
  a necessary and
  sufficient condition for $\mathcal{C}_{(1,e,s)}$ to be optimal. We only need to consider the solution of
  ${(1+x)}^{e}+{x}^{e}+1=0$ without the restriction on $\eta(x)$. In this sense,
  Our criteria   is better than that  given    in \cite[Theorem 4]{fanzhagn20}, where two systems of
  equations should be considered under some restrictions on $\eta(x)$.
\end{remark}


\begin{theorem}\label{45h-1thm}
 Let $e={5}^{m}-{5}^{m-h}+3$.
 Then $\mathcal{C}_{(1,e,s)}$ has parameters  $[{5}^{m}-1,{5}^{m}-2m-2,4]$ if $h$
 and $m$ satisfy one of  the   following conditions:

\begin{itemize}
  \item $h=0$;
  \item  $h=1$, $m \notequiv 0 \pmod9$ and $m \notequiv 0 \pmod8$.
\end{itemize}
 \end{theorem}

{\em Proof:}
Since $e\equiv 3\pmod 4$, then   $e\notin {C}_{1}$.
If $h=0$, then $\gcd(e,5^m-1)=\gcd(3,5^m-1)<p-1$, thus  $|C_e|=m$ by Lemma \ref{ce}.
If $h=1$ and $m \notequiv 0 \pmod9$,  then $\gcd(e,5^m-1)=\gcd(19,5^m-1)=1$, thus $|C_e|=m$ by Lemma \ref{ce}.

According to Theorem \ref{p5e3cor}, we need to show that
$ {(1+x)}^{e}+{x}^{e}+1=0$ has no  solution in $\mathbb{F}_{{5}^{m}}\backslash \mathbb{F}_{5}$.
Consider
\begin{eqnarray}\label{45h-1eqn1}
  ((x+1)^e+x^e+1)^{5^h}&=&\left( (x+1)^{{5}^{m}-{5}^{m-h}+3}+{x}^{{5}^{m}-{5}^{m-h}+3}+1\right)^{{5}^{h}} \nonumber \\
                       &=&\left( x+1 \right)^{4\cdot {5}^{h}-1}+{x}^{4\cdot {5}^{h}-1}+1=0. 
\end{eqnarray}

Let  $h=0$. Then (\ref{45h-1eqn1}) becomes  ${(x+1)}^{3}+{x}^{3}+1=0$. Since ${(x+1)}^{3}+{x}^{3}+1=2(x+1)(x-1)^2=0$, we have    $x=\pm 1$.

Let  $h=1$.   Then  (\ref{45h-1eqn1}) becomes  ${(x+1)}^{19}+{x}^{19}+1=0$.
Let $f_1(x)={(x+1)}^{19}+{x}^{19}+1$.
Note that
$\gcd(f_1(x),{x}^{5}-x)=(x+1)(x-1)$, $\gcd (f_1(x),{x}^{{5}^{2}}-x)=(x+1)(x-1)$, $\gcd (f_1(x),{x}^{{5}^{4}}-x)=(x+1)(x-1)$, and   $\gcd (f_1(x),{x}^{{5}^{8}}-x)={x}^{18}+3{x}^{17}+{x}^{16}+3{x}^{15}+{x}^{14}+{x}^{12}+{x}^{10}+4{x}^{8}+4{x}^{6}+4{x}^{4}+2{x}^{3}+4{x}^{2}+2x+4$.
By Lemma \ref{irrdivide},  $f_1(x)$ has 3 irreducible factors of degree 1 and 2 irreducible factors of degree 8.
In fact, $f_1(x)=2(x+1){(x-1)}^{2}g_1(x)$, where $g_1(x)=({x}^{8}+{x}^{7}+2{x}^{6}+
{x}^{5}+4{x}^{4}+4{x}^{3}+4{x}^{2}+x+3)
({x}^{8}+2{x}^{7}+3{x}^{6}+3{x}^{5}+3{x}^{4}+2{x}^{3}+4{x}^{2}+2x+2)$.
Thus, $f_1(x)$ has no root in $\mathbb{F}_{{5}^{m}}\backslash \mathbb{F}_{5}$ due to
$m \notequiv 0 \pmod8$.
\done

\begin{example}
		Let $p=5$, $m=5$, and $h=1$. Then  $e=5^m-5^{m-1}+3=2503 $.
   Let $\alpha $ be the generator  of   $\mathbb{F}_{{5}^{5}}^*$
with
$\alpha^5 + 4\alpha  + 3=0$. Then  $\mathcal{C}_{(1,e,s)}$
 has  parameters $[3124,3113,4]$ and
 generator polynomial
   $x^{11} + 4x^{10} + 4x^8 + 4x^7 + 4x^6 + 4x^5 + x^4 + 2x^3 + 3x + 1$.
\end{example}

\begin{theorem}\label{p5-3thm}
Let $p=5$ and $m$ be an odd integer.
 Let  $e=\frac{{5}^{m}-1}{2}-3$.   Then  $\mathcal{C}_{(1,e,s)}$ has parameters  $[{5}^{m}-1,{5}^{m}-2m-2,4]$.
\end{theorem}
{\em Proof:}
  Since $e\equiv 3\pmod 4$, we have
  $e\notin {C}_{1}$.
 Note that $\gcd(e,p^m-1)\leq \gcd(e,\frac{p^m-1}{2})\gcd(e,2)=\gcd(e,\frac{p^m-1}{2})=
 \gcd(3,\frac{p^m-1}{2})= 1$ due to $m$ is odd.  By Lemma \ref{ce}, $|C_e|=m$.

 According to Theorem \ref{p5e3cor}, we need to show that
$ {(1+x)}^{e}+{x}^{e}+1=0$ has no  solution in $\mathbb{F}_{{5}^{m}}\backslash \mathbb{F}_{5}$.
We consider the solution of $(x+1)^e+x^e+1=(x+1)^{\frac{5^m-1}{2}-3}+x^{\frac{5^m-1}{2}-3}+1=0 $ in the following four  cases.

Case 1), $\eta(x)=1$ and  $\eta(x+1)=1$. In this case, we get
  $(x+1)^{-3}+x^{-3}+1=0$.
Multiplying
both sides of the above equation by
$(x+1)^3x^3$, we obtain
$(x+1)^3x^3+(x+1)^3+x^3=x^6 + 3x^5 + 3x^4 + 3x^3 + 3x^2 + 3x + 1=0$.
Let $f(x)=x^6 + 3x^5 + 3x^4 + 3x^3 + 3x^2 + 3x + 1$.
Since $\gcd(f(x),x^{5}-x)=1$, $\gcd(f(x),x^{5^2}-x)=1$, and $\gcd(f(x),x^{5^3}-x)=1$,
then
$f(x)$ is irreducible
over $\F_{5}$ and  $f(x)$ has no solution in $\F_{5^m}$ if $ m\notequiv 0\pmod 6$.

Case 2), $\eta(x)=1$ and $\eta(x+1)=-1$. In this case, we get   $-(x+1)^{-3}+x^{-3}+1=0$.
Multiplying
both sides of the above equation by
$(x+1)^3x^3$, we obtain
$(x+1)^3x^3+(x+1)^3-x^3=x^6 + 3x^5 + 3x^4 + x^3 + 3x^2 + 3x + 1
=(x-1)^2(x+2)^2(x-2)^2=0$, i.e., $x=1,\pm2$.

Case 3), $\eta(x)=-1$ and  $\eta(x+1)=1$. In this case, we get
  $(x+1)^{-3}-x^{-3}+1=0$.
Multiplying
both sides of the above equation by
$(x+1)^3x^3$, we obtain
$(x+1)^3x^3-(x+1)^3+x^3=x^6 + 3x^5 + 3x^4 + x^3 + 2x^2 + 2x + 4=0$.
Let $f(x)=x^6 + 3x^5 + 3x^4 + x^3 + 2x^2 + 2x + 4$.
Since $\gcd(f(x),x^{5}-x)=1$, $\gcd(f(x),x^{5^2}-x)=1$, and $\gcd(f(x),x^{5^3}-x)=1$,
then
$f(x)$ is irreducible
over $\F_{5}$ and  $f(x)$ has no solution in $\F_{5^m}$ if $ m\notequiv 0\pmod 6$.

Case 4), $\eta(x)=-1$ and $\eta(x+1)=-1$. In this case, we get   $-(x+1)^{-3}-x^{-3}+1=0$.
Multiplying
both sides of the above equation by
$(x+1)^3x^3$, we obtain
$(x+1)^3x^3-(x+1)^3-x^3=x^6 + 3x^5 + 3x^4 + 4x^3 + 2x^2 + 2x + 4
=0$. Let $f(x)=x^6 + 3x^5 + 3x^4 + 4x^3 + 2x^2 + 2x + 4. $
Since $\gcd(f(x),x^{5}-x)=1$, $\gcd(f(x),x^{5^2}-x)=1$, and $\gcd(f(x),x^{5^3}-x)=1$,
then
$f(x)$ is irreducible
over $\F_{5}$ and  $f(x)$ has no solution in $\F_{5^m}$ if $ m\notequiv 0\pmod 6$.

Thus  $(x+1)^e+x^e+1=0 $ has no solution in $\mathbb{F}_{{5}^{m}}\backslash \mathbb{F}_{5}$.
This completes the proof.
\done

\begin{example}
Let $p=5$ and   $m=5$.  Then  $e=\frac{5^5-1}{2}-3=1559$.
   Let $\alpha $ be the generator  of   $\mathbb{F}_{{5}^{5}}^*$
with
$\alpha^5 + 4\alpha  + 3=0$. Then  $\mathcal{C}_{(1,e,s)}$
 has  parameters $[3124,3113,4]$ and
 generator polynomial
 ${x}^{11}+4{x}^{10}+2{x}^{8}+4{x}^{7}+{x}^{6}+4{x}^{5}+3{x}^{4}+{x}^{3}+3{x}^{2}+4x+1$.
\end{example}

\section{Conclusions}

 In this paper,  four classes of  optimal  $p$-ary  cyclic codes $\mathcal{C}_{(1,e,s)}$ were constructed by
 analyzing the  solutions of certain  equations over $\F_{p^m}$. It turns out that
 some previous results about  optimal quinary cyclic codes given in \cite{fanzhagn20} and \cite{tianzhang17}
  are special cases of our constructions.
Moreover, by analyzing the irreducible factors of certain polynomials over $\F_{5^m}$,
 we presented
  two classes of
 optimal quinary cyclic codes with parameters $[5^m-1,5^m-2m-2,4]$.
 At the end of this paper,
 according to our
Magma experimental data,
 we propose
 two open problems about quinary cyclic codes, it would be nice if they could be settled.

{\em Open Problem 1:}
Let  $p=5$ and  $ m $ be an odd integer. Let $e=4(5^h+1)$, where  $0\leq h\leq m-1$.
Is it true that  $\mathcal{C}_{(1,e,s)}$ has parameters  $[{5}^{m}-1,{5}^{m}-2m-2,4]$?

For $2\leq m\leq 5$,  the answer to this question is positive and confirmed by
Magma.

{\em Open Problem 2:}
Let  $p=5$ and  $ m $ be an odd integer. Let $e=5^h-2$, where  $1\leq h\leq m-1$.
Is it true that  $\mathcal{C}_{(1,e,s)}$ has parameters  $[{5}^{m}-1,{5}^{m}-2m-2,4]$?

For $2\leq m\leq 5$,  the answer to this question is positive and confirmed by
Magma.

%
%
%


\begin{thebibliography}{1}

\bibitem{carletdingit05}
C. Carlet, C. Ding, J. Yuan. Linear codes from highly nonlinear functions and their secret sharing schemes, IEEE Trans. Inf. Theory. 51 (6) (2005) 2089-2102.

\bibitem{Ding-Helleseth} C. Ding, T. Helleseth, Optimal ternary cyclic codes from monomials, IEEE Trans. Inf. Theory 59 (9) (2013) 5898-5904.

\bibitem{dinggaozhouit13}
C. Ding, Y. Gao, and Z. Zhou, Five families of three-weight ternary
cyclic codes and their duals, IEEE Trans. Inf. Theory    59 (12) (2013) 7940-7946.

\bibitem{fanlizhouffa16}
C. Fan, N. Li, Z. Zhou, A class of optimal ternary cyclic codes and their duals, Finite Fields Appl. 37 (2016) 193-202.

\bibitem{fanzhagn20}
J. Fan, Y. Zhang. Optimal quinary cyclic codes with minimum distance four, Chinese Journal of Electronics. 29 (3) (2020) 515-524.

\bibitem{huffman10}
C. Huffman, V. Pless, Fundamentals of error-correcting codes. Cambridge university press, (2010).

\bibitem{hanyanffa19}
D. Han, H. Yan, On an open problem about a class of optimal ternary cyclic codes, Finite Fields Appl. 59 (2019) 335-343.

\bibitem{lizhou15}
N. Li, Z. Zhou, T. Helleseth, On a conjecture about a class of optimal ternary cyclic codes, in:
Seventh International Workshop on Signal Design and Its Applications in Communications, 2015,
pp. 62-65.


\bibitem{liffa14}
N. Li, C. Li, T. Helleseth, C. Ding X. Tang, Optimal ternary cyclic codes with minimum distance four and five. Finite Fields Appl. 30 (2014) 100-120.


\bibitem{lidl83}
R. Lidl, H. Niederreiter, Finite Fields, Encycl. Math. Appl., vol. 20, Addison–Wesley, Reading, MA,
(1983).




\bibitem{liucaoludcc2020}
Y. Liu, X. Cao, W. Lu,
On some conjectures about optimal ternary cyclic codes. Des. Codes Cryptogr. 88(2) (2020) 297-309.






\bibitem{liuwangccds22}
K. Liu, Q. Wang, H. Yan, A class of binary cyclic codes with optimal parameters. Cryptography and Communications.  14 (3) (2022) 663-675.

\bibitem{liucao21}
Y. Liu, X. Cao, W. Lu. Two classes of new optimal ternary cyclic codes. Advances in Mathematics of Communications, 2021, doi: 10.3934/amc.2021033.

\bibitem{lizhu19}
L. Li, S. Zhu, L. Liu, Five classes of optimal ternary cyclic codes and the weight distributions of their duals. Chinese Journal of Electronics. 28 (4) (2019) 674-681.

\bibitem{liucao20}
Y. Liu, X. Cao, Four classes of optimal quinary cyclic codes. IEEE Communications Letters. 24 (7) (2020) 1387-1390.


\bibitem{storer67}
T. Storer, Cyclotomy and difference sets, Lect. Adv. Math., Markham, Chicago, IL, 1967.


\bibitem{tianzhang17}
Y. Tian, Y. Zhang, Optimal quinary cyclic codes with minimum distance four, Journal on Communication. 38 (2) (2017) 2017030:74-80.






\bibitem{wangcao22}
D. Wang, X. Cao, A family of optimal ternary cyclic codes with minimum distance five and their duals. Cryptography and Communications. 14 (1) (2022) 1-13.








\bibitem{wangwu16}
L. Wang, G. Wu, Several classes of optimal ternary cyclic codes with minimal distance four, Finite Fields Appl. 40 (2016) 126-137.

\bibitem{xiongli15}
M. Xiong, N. Li, Optimal cyclic codes with generalized Niho-type zeros and the weight distribution, IEEE Trans. Inf. Theory. 61 (9) (2015) 4914-4922.





\bibitem{xucaoccds16}
G. Xu, X. Cao, S. Xu, Optimal $p$-ary cyclic codes with minimum distance four from monomials, Cryptogr. Commun. 8 (4) (2016) 541-554.

\bibitem{yanzhouduffa18}
H. Yan, Z. Zhou, X. Du,
A family of optimal ternary cyclic codes from the Niho-type exponent. Finite Fields Appl. 54 (2018) 101-112.


\bibitem{zhaoluoffa22}
H. Zhao, R. Luo,  T. Sun. Two families of optimal ternary cyclic codes with minimal distance four. Finite Fields Appl. 79 (2022) 101995.

\bibitem{zhahuffa2021}
	Z. Zha, L. Hu, Y. Liu, X. Cao,
Further results on optimal ternary cyclic codes. Finite Fields Appl. 75 (2021) 101898.

\bibitem{zhahu20}
Z. Zha, L. Hu. New classes of optimal ternary cyclic codes with mini mum distance four, Finite Fields Appl. 64 (2020)  101671.

\bibitem{zhouding14}
Z. Zhou, C. Ding, A class of three-weight cyclic codes, Finite Fields Appl. 25 (2014) 79-93.
\bibitem{zhoudingitc13}
Z. Zhou and C. Ding, Seven classes of three-weight cyclic codes, IEEE
Trans. Commun. 61 (10) (2013)  4120-4126.


\end{thebibliography}
\end{document}